# Optomechanical manipulation with hyperbolic metasurfaces

Aliaksandra Ivinskaya,[1,*] Natalia Kostina,[1] Alexey Proskurin,[1] Mihail I. Petrov,[1]

Andrey A. Bogdanov,[1] Sergey Sukhov,[2,3] Alexey V. Krasavin,[4] Alina Karabchevsky,[5,6]

Alexander S. Shalin,[1,7] and Pavel Ginzburg[8,9]

[1]Department of Nanophotonics and Metamaterials, ITMO University, Birzhevaja line, 14, 199034 St. Petersburg, Russia
[2]CREOL, The College of Optics and Photonics, University of Central Florida, Orlando, Florida 32816, USA
[3]Kotel'nikov Institute of Radio Engineering and Electronics of Russian Academy of Sciences (Ulyanovsk branch), 48/2 Goncharov Str., 432071 Ulyanovsk, Russia
[4]Department of Physics, King's College London, Strand, London WC2R 2LS, UK
[5]Electrooptical Engineering Unit, Ben-Gurion University, Beer-Sheva, 8410501, Israel
[6]Ilse Katz Institute for Nanoscale Science & Technology, Ben-Gurion University, Beer-Sheva, 8410501, Israel
[7]Ulyanovsk State University, Lev Tolstoy Street 42, 432017 Ulyanovsk, Russia
[8]School of Electrical Engineering, Tel Aviv University, Ramat Aviv, Tel Aviv 69978, Israel
[9]Light-Matter Interaction Centre, Tel Aviv University, Tel Aviv, 69978, Israel

Auxiliary nanostructures introduce additional flexibility into optomechanical manipulation schemes. Metamaterials and metasurfaces capable to control electromagnetic interactions at the near-field regions are especially beneficial for achieving improved spatial localization of particles, reducing laser powers required for trapping, and for tailoring directivity of optical forces. Here, optical forces acting on small particles situated next to anisotropic substrates, are investigated. A special class of hyperbolic metasurfaces is considered in details and is shown to be beneficial for achieving strong optical pulling forces in a broad spectral range. Spectral decomposition of Green's functions enables identifying contributions of different interaction channels and underlines the importance of the hyperbolic dispersion regime, which plays the key role in optomechanical interactions. Homogenised model of the hyperbolic metasurface is compared to its metal-dielectric multilayer realizations and is shown to predict the optomechanical behaviour under certain conditions related to composition of the top layer of the structure and its periodicity. Optomechanical metasurfaces open a venue for future fundamental investigations and a range of practical applications, where accurate control over mechanical motion of small objects is required.

*Corresponding author: ivinskaya@tut.by

Control over mechanical motion of small particles with laser beams opened a venue for many fundamental investigations and practical applications.[1,2,3] Optical tweezers became one of the most frequently used tools in biophysical research, since they enable measuring dynamics of processes, control and monitor forces on pico-Newton level, e.g.[4,5]

Classical optical tweezers, including the extension of the concept to holographic multi-trap configurations,[6] are based on diffractive optical elements, and can provide trapping stiffness at the expense of an increased overall optical power. This limitation also applies to trapping with structured and superoscilatory beams (e.g.[7]). A promising solution for achieving improved localization and the highest possible stiffness within the trap is to introduce auxiliary nanophotonic or plasmonic structures that enable operation with nano-confined near fields. Plasmonic tweezers, for example, utilize localized resonances of noble metal nanoantennas and provide significant improvement in trap stiffness and spatial localization of trapped objects.[8] Optical manipulation with other auxiliary tools, e.g. endoscopic techniques,[9] nano-apertures,[10] nanoplate mirrors[11] and integrated photonic devices[12,13] was also demonstrated. A special attention was paid to particles trapping next to surfaces, since a typical experimental layout of a fluid cell may include substrates of different kinds, e.g.[14,15] and others. Furthermore, controllable transport over surfaces enables a range of particles sorting applications.[13] Here, additional advantages of carefully designed surfaces in application to optomechanical manipulations will be investigated.

Introduction of additional flexibilities and degrees of freedom into manipulation schemes is required for achieving ultimate optomechanical control over particles' motion. Tractor beams are a vivid example of a peculiar phenomenon, where a particle moves against the global beam propagation direction towards a light source. Structured illumination was initially used for demonstration of this effect.[16–18] Remarkably, unstructured light, e.g., plane wave, can result in pulling force on particles, situated above planar substrates.[19,20] Besides a special case, where beads were partially immersed into a media,[19] more practical configuration with a particle above the substrate was considered in[20] and pulling forces, owing to unidirectional excitation of surface plasmons, were predicted. Other recently reported phenomena of the optical forces, tailored with planar surfaces, include the force enhancement with Bloch modes.[21,22] An extended review on optical forces acting on a particle above a substrate is given in the Supporting Information A.

Another approach to tailoring optomechanical interactions is to utilize the concept of metamaterials, which allows achieving control over propagation of light via subwavelength structuring of unit cells.[23] A special class of metamaterials is a hyperbolic medium,[24] which enhances the density of electromagnetic states and, as the result, enables controlling efficiency of scattering channels. Hyperbolic metamaterials, based on vertically aligned metal nanorods, were assessed as a platform for flexible optomechanical manipulation.[25,26] While a set of peculiar effects, including tractor beams, were predicted by employing homogenization approach,[26] consideration of near-field interactions within actual unit cells of the nanorod arrays require more detailed numerical analysis.[25] From a practical standpoint, optomechanical manipulation inside metamaterials may face additional challenges and undesirable constraints. On the other hand, metasurfaces (e.g.[27,28,29]) (in particular, hyperbolic) can share several advantages: delivering high density of electromagnetic states for tailoring scattering properties and providing physical access to manipulated objects. Hyperbolic metasurfaces have already demonstrated superior characteristics in controlling surfaces waves and scattering, e.g.[30,31] Here, optomechanical interactions over hyperbolic metasurfaces will be analysed and a set of new effects will be predicted. In particular, tailoring electromagnetic interactions with a particle by opening high density of states scattering channels will be shown to deliver puling forces in a broad spectral range owing to the hyperbolic nature of wave dispersion in the metamaterial substrate. This effect also makes the distinctive differences between this new approach to previously reported investigations (e.g.[20]).

The proposed configuration, subject to the subsequent analysis, is depicted in Fig. 1. A small dielectric bead is situated in a close proximity of layered metal-dielectric periodic substrate, which supports hyperbolic regime of dispersion of bulk modes. The structure is illuminated with a plane wave, which excites both surface plasmon and volumetric hyperbolic waves, owing to the scattering from the particle. The interplay between resonant and nonresonant contributions of different types of electromagnetic modes gives rise to strong pulling forces, which will be analysed hereafter. Green's function formalism enables identifying contributions of different scattering channels to the overall optical force and, as a result, to develop clear design rules for tailoring optomechanical interactions.

The manuscript is organized as follows: general Green's function formalism in application to optical forces will be revised first and then applied to the case of the homogeneous hyperbolic metasurface. Conditions for achieving optical pulling forces will be derived and contributions of surface and volumetric waves will be identified. Analysis of a realistic layered structure will be performed and additional peculiar optomechanical effects will be discussed. Details of mathematical analysis and several optimization routines appear in the Supporting Information.

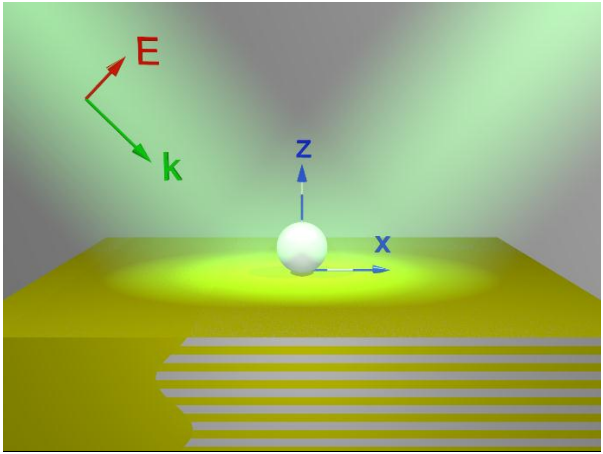

**Figure 1.** Optomechanically manipulated subwavelength dielectric particle next to anisotropic (hyperbolic) metasurface, illuminated with an obliquely incident plane wave. The origin of Cartesian coordinates is the point at the substrate-air interface below the geometric centre of the particle.

**Dipole particle over anisotropic substrate – General Formalism**

Optical forces, acting on a small subwavelength particle, obeying dipolar approximation, can be expressed by[32] $F_j = \frac{1}{2} \mathrm{Re}(\alpha \mathbf{E}^{tot} \cdot \partial_j \mathbf{E}^{tot*})$, where $\mathbf{E}^{tot}$ is a local field at the particle location, $\alpha$ is particle's complex polarizability taking into account radiation reaction, and subscript indexes stay for directions in the Cartesian coordinate system.[33] If a particle is situated next to a photonic structure, self-consistent electromagnetic field contains re-scattering contributions between all the constitutive elements. In the case of a small particle next to a flat surface, transversal optical force can be calculated self-consistently via:[20]

$$F_x = \frac{1}{2} \mathrm{Re}(\alpha \mathbf{E}^{tot} \cdot \partial_x \mathbf{E}^{0*}) + |\alpha|^2 \, \omega^2 \mu_0 \mu_1 \, \mathrm{Im}(\mathrm{E}_x^{tot} \mathrm{E}_z^{tot*}) \, \mathrm{Im}(\partial_x \, \mathrm{G}_{xz}), \tag{1}$$

where $G_{xz}$ is a component of the Green's function tensor $\ddot{\mathbf{G}}$ of a dipolar scatterer above a substrate, $\mathbf{E}^0$ is the electrical field of a wave incident and reflected from the substrate (calculated without a particle), and $\mu_0$ and $\mu_1$ (=1 in subsequent analysis) are permeabilities of vacuum and upper half-space. Green's function appears in both expressions for the total field $\mathbf{E}^{tot}$ and the force $\mathbf{F}$ (see Supporting Information B for details). Both summands in Eq. 1 influence transversal optical force $F_x$, however, for realistic materials the first term is positive and comparable with free-space force magnitude, while the second one might take large values, resulting from Green's function component $(\partial_x G_{xz})$ and excitation conditions, given by $\mathrm{Im}(E_x^{tot} E_z^{tot*})$. The product $\mathrm{Im}(E_x^{tot} E_z^{tot*})\,\mathrm{Im}(\partial_x G_{xz})$ with an appropriate sign might change the overall balance and result in an attracting force (see Supporting Information C).

Green's function encapsulates the response from the auxiliary layered structure, which manifest itself via reflection coefficients $r_s$ and $r_p$ for $s$ and $p$ polarizations. Here, the whole set of the plane waves, including propagating and evanescent ones, should be taken into account. Evanescent components are generated by the particle, which are near-field coupled to the surface. In the case of a hyperbolic substrate, both surface and bulk waves can be supported (note, that hyperbolic metamaterials convert evanescent fields into propagating ones owing to the peculiar law of dispersion). The excitation efficiency of different modes depends on density of states (DOS); in the case of surface plasmons it has strongly resonant behaviour, while hyperbolic bulk modes have high DOS over a broad spectral range. Those properties enable separating different electromagnetic channels via the Fourier spectrum representation of the Green's function and, as a result, identify the main contributing mechanism for achieving strong opto-mechanical interactions.

It is worth noting that Green's functions in flat layer geometries (one among possible ways to construct an artificial hyperbolic metamaterial) has analytical expressions, allowing direct comparison between performances of homogenized media and its possible realization. From the applied standpoint, fabrication of nanometre thin layers is possible with atomic layer deposition techniques, e.g.,[34,35,36] which makes the theoretically developed concept to be attractive from the experimental perspectives.

**Optical attraction, mediated by hyperbolic substrates**

The impact of homogenized anisotropic substrates on optical forces will be studied next. Figure 2(a),(b) shows transversal force $F_x$ acting on a small dielectric particle above a uniaxial crystal as calculated with Eq. 1. The force map is built as a function of real parts of dielectric permittivities along the two main axes $\varepsilon_x = \varepsilon'_x + i\varepsilon''_x$ and $\varepsilon_z = \varepsilon'_z + i\varepsilon''_z$ of the crystal. Regions of positive and negative values of the force are displayed with grey and colour legends respectively. Each quadrant (marked on panel (a)) on the force map corresponds to a different regime of the substrate's dispersion. The first quadrant shows the normal elliptical regime of dispersion, which does not support any surface waves and high DOS volume waves. As a result, an attraction force cannot be achieved. The third quadrant, where both components of the tensor are negative, corresponds to the regime of surface plasmon modes, while the bulk substrate does not support any propagation of waves within its volume. In this case a strong attraction can be achieved owing to the surface mode, as was previously demonstrated,[20] owing to unidirectional excitation of plasmons. Clear resonant plasmonic branch is seen on the force map. The width of the force resonance becomes wider with the increase of the material loss (panel (b)). Quadrants II and IV support the regime, when both surface plasmon and hyperbolic bulk modes can be supported. Remarkably, the main difference in the force behaviour (no attraction at the forth quadrant) emerges owing to the orientation of the polarization of the incident field with respect to the negative component of permittivity tensor. This behaviour can be retrieved by analyzing the balance between $\mathrm{Im}(\partial_x G_{xz})$ and $\mathrm{Im}(E_x E_z^*)$, see Supplementary Fig. S1 and the corresponding discussion.

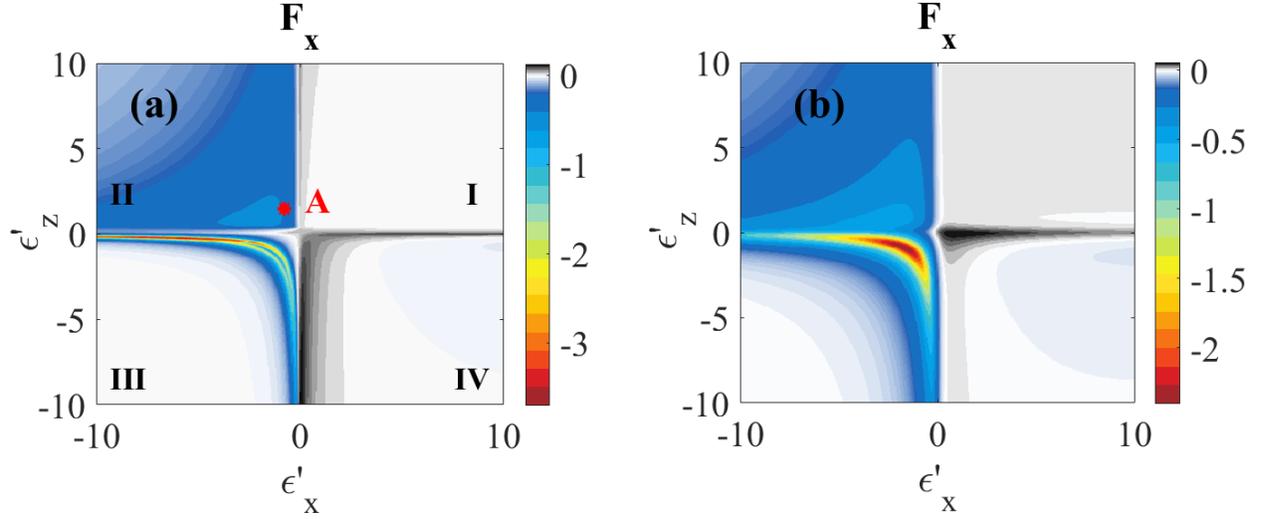

**Figure 2.** Colour maps of optical forces (in fN mW$^{-1}$μm$^2$) acting on a particle ($\varepsilon = 3$, $R = 15$ nm, $z = 15$ nm) above an anisotropic material, as the function of real parts of the substrate tensor parameters $\varepsilon'_x$, $\varepsilon'_z$. Material losses are set to (a) $\varepsilon''_x$, $\varepsilon''_z = 0.05$, (b) $\varepsilon''_x$, $\varepsilon''_z = 0.3$. Plane wave is incident at 35$^{\circ}$ in respect to the normal, $\lambda = 450$ nm. Balances between the force components at the regime, marked by point A, will be analysed in Fig. 3.

While the regime of the second quadrant supports the surface waves, a clear plasmonic branch in the force map is not observed. The balance between different contributions will be analysed next - it is worth re-emphasizing that hyperbolic materials can support both surface and bulk waves. Volumetric waves with hyperbolic dispersion have non-resonant features, whereas existence of surface modes can be identified via observation of poles in the reflection coefficient. Those coefficients for the case of the uniaxial crystal with an optical axis, pointing perpendicular to the interface, are given by:[37,38]

$$r_p = \frac{\varepsilon_x k_{1z} - \varepsilon_1 k_{2z}^p}{\varepsilon_x k_{1z} + \varepsilon_1 k_{2z}^p}, \quad k_{2z}^p = \left(\varepsilon_x k_0^2 - \frac{\varepsilon_x}{\varepsilon_z} k_\rho^2\right)^{0.5}, \quad k_{1z} = \left(\varepsilon_1 k_0^2 - k_\rho^2\right)^{0.5}, \quad (2)$$

where $k_0$ is a wave vector of incident radiation in vacuum, $\varepsilon_1$ and $k_1 = \sqrt{\varepsilon_1} k_0$ are the permittivity and the wave vector in the upper half-space (air is considered hereafter) with components $k_x$, $k_y$, $k_{1z}$ and $k_\rho = (k_x^2 + k_y^2)^{0.5}$ is a transversal wave vector. The branch of the square root solution in Eq.

2 should be chosen with imaginary part of $k_{2z}^p$ (wave vector in the substrate) positively defined for a wave to decay away from the interface. At the same time, for idealized lossless hyperbolic materials, having different signs of $\varepsilon_x', \varepsilon_z'$, wave vector $k_{2z}^p$ acquires real part for $k_\rho = k_{cr} \geq k_1$. It implies that at this condition evanescent waves scattered by the particle transform into propagating volume modes inside the hyperbolic substrate.[39] On the other hand, the pole in the reflection coefficient (Eq. 2) gives the dispersion of the surface plasmon propagating over an anisotropic substrate:

$$k_\rho^{pl} = k_0 \left( \frac{(\varepsilon_1 - \varepsilon_x)\varepsilon_1\varepsilon_z}{\varepsilon_1^2 - \varepsilon_x\varepsilon_z} \right)^{0.5}. \tag{3}$$

Surface plasmon resonance condition is fulfilled once the real part of the denominator of in Eq. 3 goes to zero.

In order to find the contributions of surface and volumetric waves to the pulling force, the integral representation of Green's function can be used. At the close proximity of a particle to the substrate ($z \ll \lambda$), several contributions to the derivative $\partial_x G_{xz}$ can be identified:

$$\partial_x G_{xz} = \frac{1}{8\pi k_1^2} \int_0^\infty r_p k_\rho^3 \, \mathrm{e}^{2ik_{1z}z} \, \mathrm{d}k_\rho = \frac{1}{8\pi k_1^2} \{ \int_0^{k_1} + \int_{k_1}^{k_{cr}} + \int_{k_{cr}}^\infty \} \, r_p k_\rho^3 \, \mathrm{e}^{2ik_{1z}z} \, \mathrm{d}k_\rho = g^{pr} + g^{pl} + g^{hb}. \tag{4}$$

This splitting of integral into $g^{pr}$, $g^{pl}$, $g^{hb}$ summands underlines the contribution of the free space propagating (in the air superstrate), plasmon and hyperbolic modes. Linking those parts of the integral with the beforehand mentioned physical interpretation can be done under several approximations. First, the particle should be located in the close proximity of the substrate in order to excite both plasmon and volume modes efficiently. Second, the plasmonic mode should have a narrow pole in the reflection coefficient, which is perfectly satisfied in the case of the low loss hyperbolic material.

Three spectral intervals, underlined in Eq. 4, can be identified by observing the imaginary part of the reflection coefficient $r_p$ , Fig. 3(a) (s-wave reflects similar to pure metal case and $r_s$ does not contribute to $\partial_x G_{xz}$). As it is shown by Eq. S8 in Supporting Information A, $\mathrm{Im}(\partial_x G_{xz})$ is primarily defined by $\mathrm{Im}(r_p)$ and the proposed separation of different modes contributions to the optical force given by Eq. 4 and is justified.[39] In the hyperbolic regime, the reflection coefficient is virtually constant at larger $k_\rho$ and using $r_p^\infty = r_p(k_\rho / k_1 \to \infty)$ for evaluation of $g^{hb}$ (one of the summands in Eq. 4) is a rather good approximation. With $r_p^\infty = (\sqrt{\varepsilon_x \varepsilon_z} - \varepsilon_1) / (\sqrt{\varepsilon_x \varepsilon_z} + \varepsilon_1)$ the term $g^{hb}$ can be analytically integrated to the simple form:

$$g^{hb,ap} = \frac{3 r_p^\infty}{4\pi k_1^2 (2 k_1 z)^4} \quad .\tag{5}$$

Note, that the exact value of $k_{cr}$ has only a minor contribution to the integration of $g^{hb,ap}$ as long as $k_{cr} z$ is small (this is the manifestation of the nonresonant nature of high DOS in homogeneous hyperbolic metamaterials). $g^{hb,ap}$ is symmetric with respect to $\varepsilon_x, \varepsilon_z$ as follows from the expression for $r_p^\infty$. Fig. 3(b) demonstrates the plot of the term $g^{hb,ap}$ together with actual value of $\partial_x G_{xz}$. Here either $\varepsilon_x'$ or $\varepsilon_z'$ is fixed to -0.5 while another component is varied and $\partial_x G_{xz}$ is evaluated. Note, that $g^{hb,ap}$ has the same value if $\varepsilon_x'$ and $\varepsilon_z'$ interchange their roles, while $\partial_x G_{xz}$ demonstrates a slight difference in the real part. It is evident that $g^{hb,ap}$ strongly prevails the overall value of $\partial_x G_{xz}$ and all other contributions (see Eq. 4) can be neglected. As a result, pulling forces, observed in quadrant II (Fig. 2), *mainly arise due to volume hyperbolic modes*. Finally, it is worth noting, that optical forces are not symmetric in respect with interchanging the roles of $\varepsilon_x'$ and $\varepsilon_z'$, since the polarization of the excitation and direction of propagation affects differently the value of $\mathrm{Im}(E_x^{tot} E_z^{tot*})$ (Eq. 1) - see the Supporting Information C.

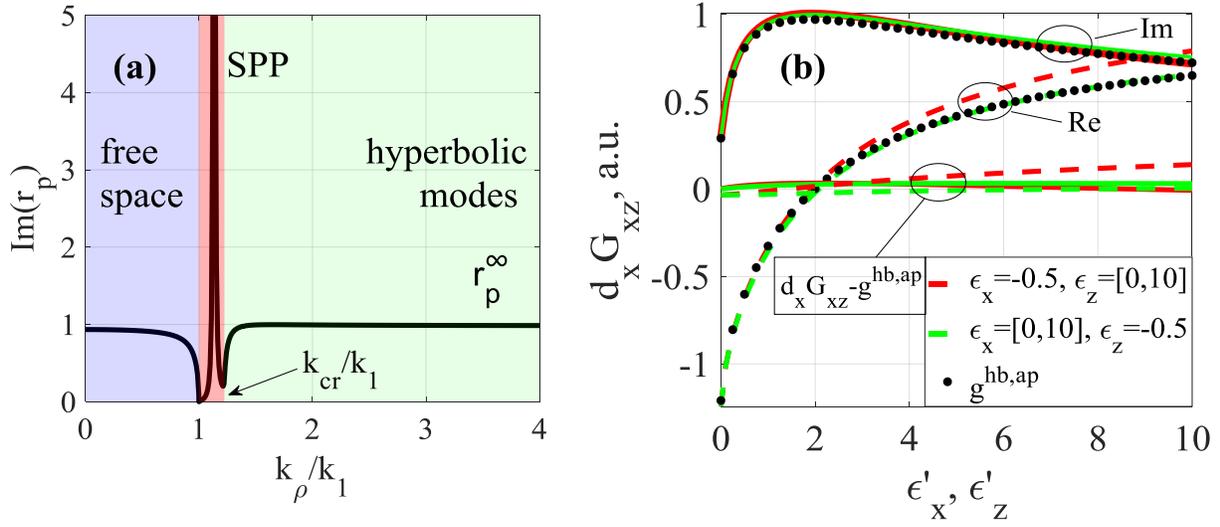

**Figure 3.** (a) Reflection coefficient for hyperbolic material as a function of transversal wave vector $k_\rho$. The parameters of the system correspond to point A from Fig. 2(a) with $\varepsilon_x' = -0.5, \varepsilon_z' = 1.1, \varepsilon_x'', \varepsilon_z'' = 0.01$. (b) Green's function derivative $\partial_x G_{xz}$ (both real and imaginary parts) above the anisotropic substrate at coordinate $z = 15$ nm. Two cases are plotted: $\varepsilon_x' = -0.5, \varepsilon_z' \in [0,10]$ (red lines) and, vice versa, $\varepsilon_x' \in [0,10], \varepsilon_z' = -0.5$ (green lines). Black dots correspond to approximate value $g^{hb,ap}$ (Eq. 5). Imaginary part of the permittivity was taken to be $\varepsilon_x'', \varepsilon_z'' = 0.05$.

## Multilayer substrates for negative transversal force

Metal-dielectric multilayers are one among probable realizations of hyperbolic metamaterials.[35] The goal of the subsequent studies is to compare performance of realistic composites with their homogenised counterparts, investigated in the previous section. Silver layers interchanging with a transparent dielectric (e.g. made of porous polymer, refractive index $n = 1.05$) will be studied next. Effective material properties can be attributed to those composites if the periodicity of the structure and thicknesses of layers are significantly smaller than the wavelength. In this case, the effective dielectric tensor is given by $\varepsilon_x = \varepsilon_1 f_1 + \varepsilon_2 (1-f_1), \varepsilon_z = \left(f_1 \varepsilon_1^{-1} + (1-f_1)\varepsilon_2^{-1}\right)^{-1}$ where $\varepsilon_1, \varepsilon_2$ are permittivities of metal and dielectric, $f_1 = d/P$ is filling fraction of metal of thickness $d$ in the unit cell of period $P$.[39] While this homogenization approach works extremely well in the case of a small dielectric contrast between the constitutive materials, it might face several limitations, if plasmonic layers are alternating with positive epsilon dielectrics.[34] Spatial dispersion effects (e.g.[40]), fast-decaying evanescent waves next to metamaterial-air interfaces,[41] local near-field corrections (e.g.[39]) and others should be taken into account,[25,26] depending on a specific phenomenon under consideration. Hereafter, homogenization approach to layered metamaterials is assessed in application to optical forces (see also Supporting Information D).

The exact reflection coefficients for semi-infinite multilayer are obtained according to the approach developed in.[39] The material composition of the first layer (either metal or dielectric) has a crucial impact on the pulling force. This effect is solely related to the realization of the metamaterial and, of course, cannot be observed in the studies of the homogenized substrate. Those three different scenarios, first metal layer, first dielectric layer, and homogeneous substrate are considered hereafter. Fig. 4(a) demonstrates the value of the pulling force (negative value = pulling), as the function of wavelength of the incident radiation (the silver filling factor is fixed at $f_1 = 0.2$). Corresponding homogenized substrate has effective parameters, corresponding to point A in Fig. 2(a). Results of a finite-element method (FEM) simulation, performed with Comsol Multiphysics, are shown with black dots. The values of the force for the homogenized case typically lie in between those of multilayer realizations with either dielectric (blue lines) or metal (red lines) top layer. If metal is set as an outer layer, maximum force amplification is achieved. At the longer wavelengths (infrared) regimes of metal or dielectric top layer can give an order of magnitude difference in force values (see Supporting Fig. S3). As it is expected, multilayers with smaller period ($P$ in the captions) (closely resembling the homogenized description) have broadband spectra, while larger periods give sharper features and a slight blue shift of the resonant frequency. The homogenized model perfectly predicts the behaviour of the multilayer composite, if the period of the latter is as small as 0.5-1 nm, which seems to be unrealistic with the current fabrication technologies. Furthermore, in order to underline features, provided by multilayers versus homogeneous substrate, the particle was chosen to touch the outer layer of the composite. As the distance to substrate is increased, the discrepancies between two approaches become minor.[42] In order to verify the validity of the developed theoretical approach, full wave finite element numerical simulation was performed (black dots). 30 nm particle above homogeneous substrate was investigated. Qualitatively good agreement between the numerical and analytical results was achieved. The discrepancies are attributed to the finite volume of the particle, which was fully accounted in the case of FEM simulation and was considered in the dipolar approximation in the analytical theory.

Additional optimization over the layers' parameters in order to achieve maximal pulling force can be performed and is discussed in the Supporting Information D and E. It should be noted that many experimental investigations of optical forces are done with particles, dissolved in liquid solutions. In such environments, optically denser particles can be used in order to achieve sufficient force amplitude, see the Supporting Information F.

To get a deeper insight into the differences between homogenised and layered cases, the imaginary part of the reflection coefficient was plotted, as the function of the transversal wave vector of incident wave - Fig. 4(b). Metal-dielectric stacks with different periodicity, but with the same filling factors (homogenized parameters are supposed to be the same), were investigated. Multilayer substrates feature strongly non-uniform dependence of $\mathrm{Im}\left(r_p(k_\rho)\right)$. The lowest reflectivity of evanescent waves is obtained in the case, where the top layer is the dielectric. As a result, the lowest pulling force values are obtained (see panel (a)). As metal layer is placed on the top, situation changes drastically. Sharp peak in $r_p$ at $k_\rho$ slightly exceeding $k_1$ is linked with the existence of the long-range plasmon (LR-SPP). Furthermore, at $k_\rho \gg k_1$ reflection coefficient obtains significant values in the allowed zone of the periodic structure. Here the reflection coefficient exceeds $r_p^\infty$ (characteristic to homogenous hyperbolic material at large $k_\rho$) due to the short-range plasmon (SR-SPP).[39] Convolution of such $r_p(k_\rho)$, enhanced by short-range plasmon, with exponentially decaying term in the integrand of Eq. 4 defines features, characteristic to different periods of the unit cell.

Concluding this part of the investigation, both volume modes and SR-SPP, linked with $\mathrm{Im}\left(r_p(k_\rho)\right)$, are crucial for optical forces. Increasing $\mathrm{Im}\left(r_p(k_\rho)\right)$ at the region of evanescent waves can lead to the enhancement of the puling force. As for SPP resonance at the bulk metal interface, considered previously in,[20] it is also characterized by $\mathrm{Im}\left(r_p(k_\rho)\right)$ increase for a broad range of evanescent $k_\rho$. However, in this case the effect is associated with a sharp spectral resonance (plasmonic branch). For low-loss metals it results in strong pulling forces, e.g., for Ag substrate tractor beam takes place at wavelength around 340 nm, corresponding to SPP resonance.

Besides narrow-band spectral dependence, the amplitude of pulling force will drop significantly with an increase of absorption in metal, what complicates the experimental applications of SPP resonance-assisted tractor beam in the visible range. Note, that different physical trends are expected for optical forces due to different kind of modes at the substrate interface and within the bulk. In particular, spectral features and loss dependence of volumetric modes can be tailored and subsequently employed for achieving significant values of pulling forces.

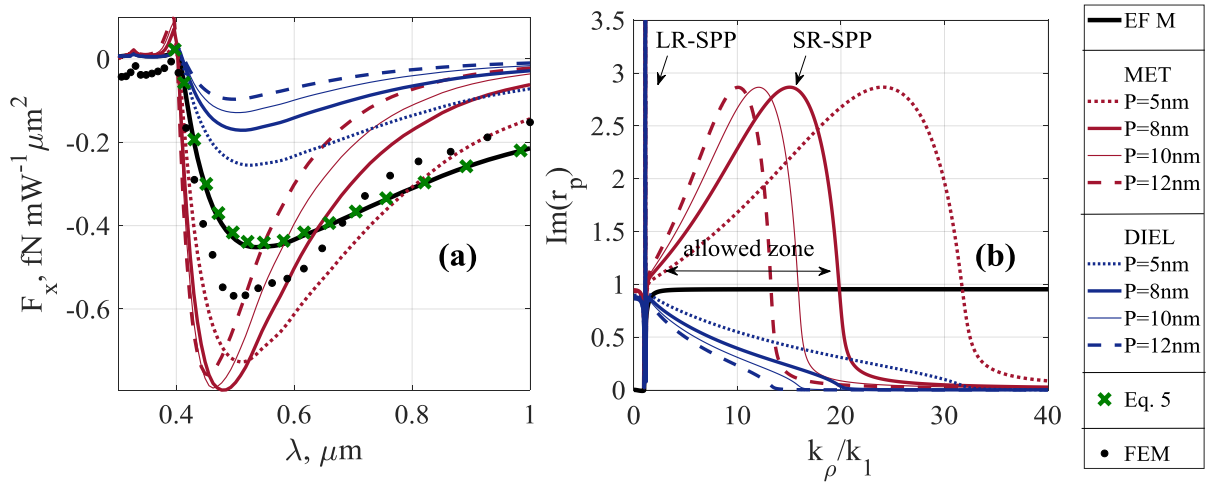

**Figure 4.** (a) Transversal force $F_x$ acting on a particle, as a function of the wavelength. (b) Imaginary part of the reflection coefficient for semi-infinite Ag-polymer multilayer composite as the function of the transverse wave vector. Red (blue) lines correspond to the case, when the top layer, interfacing the air, is metal (dielectric). Black line is obtained within the effective medium approach; green crosses correspond to approximation given by Eq. 5; black dots correspond to numerical simulation of a particle a distance $z = 16.5$ nm above anisotropic substrate. A set of multilayer periods $P$ is specified in the legend. Other parameters are: $R = 15$ nm, $\varepsilon = 3$, $z = 15$ nm, $f_1 = 0.2$, the structure is illuminated by a plane wave incident at 35°.

While the beforehand studies of the forces were performed on the configurations, where the metal was chosen to be silver, other plasmonic materials are of a potential interest (especially with the development of other promising alternatives[43]). Hereafter, Au (3-5nm) and Ag (1.5-3nm) thick layers will be compared with each other and with semi-infinite substrates, made of a solid metal (traditional plasmonic substrate). Fig. 5 summarizes the results. The pulling force at pure metal substrate emerges solely owing to the interaction with surface plasmon polaritons at the resonance. Bulk silver, having smaller optical losses than gold, provides very strong optical force, while gold substrate gives very low force amplitude. Single isolated Ag layer can also provide a pulling force due to SR-SPP, however, this effect is relatively weak and is not presented here. For both Au and Ag multilayers the force spectra are red-shifted with respect to the bulk substrates. In addition, the force bandwidths are significantly wider for the layers. Remarkably, in the multilayer hyperbolic case, pulling forces are much stronger than for a bulk Au substrate and less affected by material loss – both gold and silver layers show similar amplitudes. Broadband force enhancement is achieved with multilayer substrates in the whole visible range from 400 to 800 nm. Unlike purely plasmonic case, hyperbolic modes are less affected by losses. It is worth noting that virtually any metal provides noticeable negative force (Supporting Information E) and is suitable for optomechanical applications.

Insert in Fig. 5 shows a FEM simulation of magnetic field scattered by a particle (excluding background field given by the incident and reflected plane waves) for the homogeneous hyperbolic material. Highly confined high-k volume modes are seen in the substrate and are typical to hyperbolic metamaterial (e.g.[44]). The asymmetric scattering pattern within the substrate visualizes the puling force effect and the impact of the metamaterial. The substrate opens the high DOS channels and allows tailoring forces by breaking the symmetry of the momentum, carried out by the scattered waves. Furthermore, high-k waves carry more momentum in comparison to free space counterparts and, as the result, recoil force values increase.

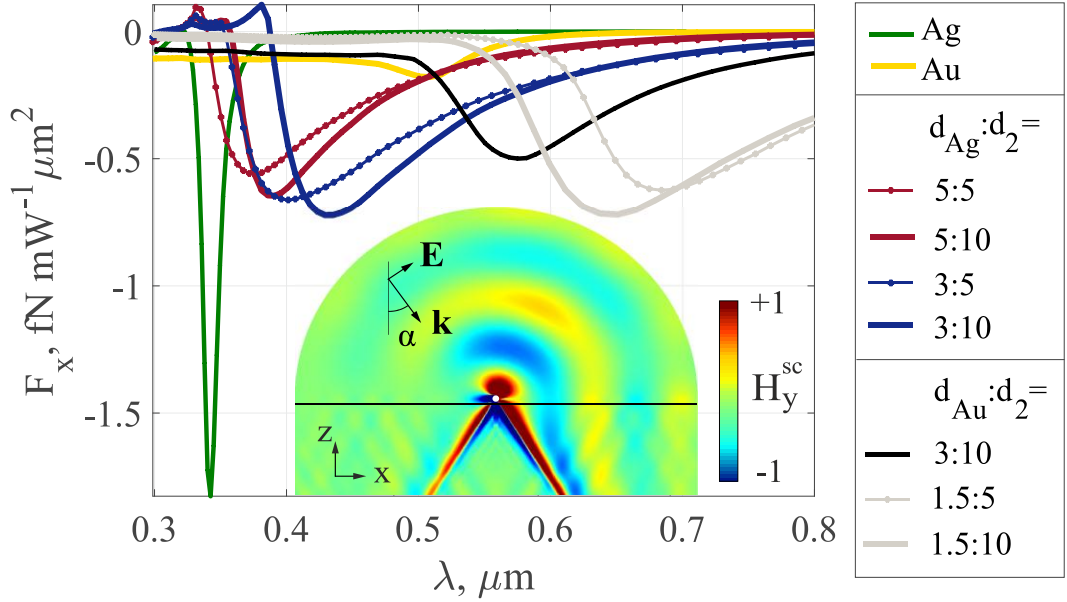

**Figure 5.** Transversal force $F_x$ acting on a particle ($R$ = 15 nm, $\varepsilon$ = 3, $z$ = 15 nm) for different types of substrates: silver (green line), gold (yellow line), Ag-polymer multilayer (red and blue lines) and Au-polymer multilayer (black and grey lines). In legend $d_{Ag}$, $d_{Au}$, $d_2$ are specified in nm and stand for Ag, Au and polymer thicknesses. Multilayer is terminated with metal; plane wave is incident at 35 º. The inset shows magnetic component $H_y^{sc}$ of the field scattered by the particle above the homogeneous hyperbolic substrate (for the effective model of Ag-polymer multilayer $f_{Ag}$ =0.2 at $\lambda$ = 450 nm), $z$ = 16 nm, $\alpha$ = 35 º.

Summarizing, multi-layered metamaterials can be used as effective substrates for achieving negative optical forces, which, compared to the plasmonic substrates, provide a set of new features: 1) broadband pulling force (instead of narrowband response at the surface plasmon resonance or a weak response at SR-SPP), 2) tuning the position of the force spectra by appropriate design of the multilayer structure, 3) large negative force for multilayers with lossy metals.

## Conclusions

Optical forces mediated by hyperbolic metamaterials interfacing dielectric homogeneous space (e.g. air or water, in the case of fluidic applications) were theoretically investigated. Strong optical pulling forces were obtained over the broad spectral range and this new effect is attributed to the hyperbolic type of dispersion of electromagnetic modes. Analytical formalism of the phenomenon was developed and it relies on the self-consistent expressions for the total field at the particle location, given with the help of electromagnetic Green's function. Spectral decomposition of the Green's function enabled identifying contributions of different interaction channels into the total optical force. In particular, long-range surface plasmon polaritons having sharp resonance in k-space were separated from the hyperbolic bulk modes for the case of homogeneous hyperbolic material. Hyperbolic modes, having high density of electromagnetic states over a broad spectral range, are the preferable channel for scattering. It was shown that the hyperbolicity of the substrate plays the key role in delivering the overall effect in the broad spectral range. However, multilayered realization of the metamaterial, when the top material interfacing the air is metal, introduces an important contribution to the value of the pulling force. The broadband property and tunability of the hyperbolic metasurface have an additional key advantage over a single metal layer – optical attraction can be achieved at the infrared spectral range, which is highly important to many optomechanical applications. Metasurfaces, designed to control near-field interactions, open a venue for flexible optomechanical manipulation schemes. While the effect of optical attraction was demonstrated here, other important fundamental phenomena and applications, such as optical binding, sorting, trapping, to name just few, can be enabled by carefully designed optomechanical metasurfaces.

## Author Contributions

AI designed the model and the computational framework and analysed the data. AI and PG wrote the manuscript. AB and AP performed finite element simulations. PG and AS supervised the work. All authors contributed to the research activities.

## Funding Sources

The authors gratefully acknowledge the financial support provided by Russian Fund for Basic Research (Project No. 18-02-00414, 18-52-00005); Ministry of Education and Science of the Russian Federation (Grant No. 3.4982.2017/6.7); The force calculations were partially supported by Russian Science Foundation (Grant No. 18-72-10127). AK acknowledges the support of the Israeli Ministry of Trade and Labor-Kamin Program, Grant. No. 62045.

# Optomechanical manipulation with hyperbolic metasurfaces

Aliaksandra Ivinskaya,[1,*] Natalia Kostina,[1] Alexey Proskurin,[1] Mihail I. Petrov,[1]

Andrey A. Bogdanov,[1,2] Sergey Sukhov,[3,4] Alexey V. Krasavin,[5] Alina Karabchevsky,[6,7]

Alexander S. Shalin,[1,8] and Pavel Ginzburg[9,10]

[1]Department of Nanophotonics and Metamaterials, ITMO University, Birzhevaja line, 14, 199034 St. Petersburg, Russia
[2]Ioffe Institute, St. Petersburg 194021, Russia
[3]CREOL, The College of Optics and Photonics, University of Central Florida, Orlando, Florida 32816, USA
[4]Kotel'nikov Institute of Radio Engineering and Electronics of Russian Academy of Sciences (Ulyanovsk branch), 48/2 Goncharov Str., 432071 Ulyanovsk, Russia
[5]Department of Physics, King's College London, Strand, London WC2R 2LS, UK
[6]Electrooptical Engineering Unit, Ben-Gurion University, Beer-Sheva, 8410501, Israel
[7]Ilse Katz Institute for Nanoscale Science & Technology, Ben-Gurion University, Beer-Sheva, 8410501, Israel
[8]Ulyanovsk State University, Lev Tolstoy Street 42, 432017 Ulyanovsk, Russia
[9]School of Electrical Engineering, Tel Aviv University, Ramat Aviv, Tel Aviv 69978, Israel
[10]Light-Matter Interaction Centre, Tel Aviv University, Tel Aviv, 69978, Israel

# Table of Content

Pages: S1-S11

Tables: S1

Equations: S1-S10

Figures: S1-S5

### A. Literature review

For a particle placed over the substrate or waveguide, direction of horizontal and vertical forces might change essentially depending on the illumination scheme, particle and substrate composition as well as modes excited in the particle, bound at the interface or supported by the waveguide [1-15], see Table 1.

In the last decade a significant research effort was devoted to the studies of the optical properties of isotropic substrates. For example, Kretschmann configuration (plot (1) in Table S1), when a source is positioned below a substrate, is well studied. For dielectric substrates, for incidence angles larger than the angle of total internal reflection (TIR) the field is purely evanescent in the upper half-space. On the contrary, at smaller angles propagating harmonics are present above the dielectric substrate. In both cases for dielectric particles vertical force $F_z$ (perpendicular to the dielectric substrate) can be both repulsive or attractive [3]. For Ag particles under TIR condition repulsion or attraction to the substrate is defined by the change of the sign of particle polarizability which changes at the configuration resonance [4]. Multilayer substrates and Ag-coated dielectric beads are considered in [5]. Horizontal force (along the substrate) was shown to point in the direction of light propagation for both dielectric and Ag particles [4,5].

It was shown that a radiating dipole can be repulsed from a substrate (levitation) [6–8], plot (2) in Table S1.

In a case of a waveguide substrate (plot (3) in Table S1) [9–12], evanescent fields can cause both vertical repulsion and attraction of large particles to waveguide [9]. In horizontal direction dielectric beads can be propelled [9,10] or, vice versa, move opposite to the power flow [13] depending on the mutual orientation of phase and group velocities of a particular waveguide mode. When an illumination is incident on a substrate from the top (plot (4) in Table S1), the field is a combination of evanescent and propagating waves, and we can expect a rich variety of phenomena. Indeed, the optical force can take an arbitrary direction for large Au particles [13], while for small dielectric particles the pulling force could be achieved due to surface plasmons [14]. Enhancement of vertical forces is studied in [15]. We extend the studies of horizontal forces by exploring uniaxial anisotropic substrates (optical axis is perpendicular to interface) supporting extraordinary hyperbolic modes. Considering nanoparticles without optical resonances we can attribute optomechanical effects to the illumination scheme and modes of the substrate only.

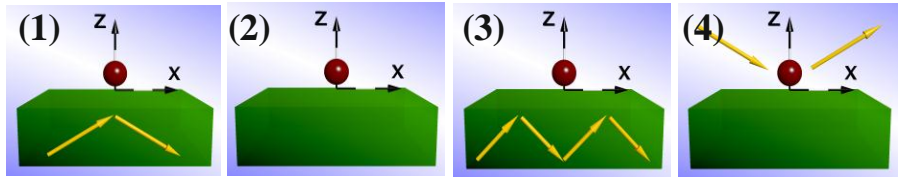

| setup | | (1) | (2) | (3) big/sm. part. | (4) |
|---|---|---|---|---|---|
| Fx | push | + | | +/+ | + |
| | pull | | | +/+ | + |
| Fz | attr. | + | + | +/+ | + |
| | rep. | + | + | +/- | + |

**Table S1.** Summary of the optical force direction for particles above the substrate. (1) Kretshmann configuration [3–5]; (2) a radiating dipole can be attracted or repulsed from the substrate even at the absence of external field [6–8]; (3) a particle above a waveguide [9–12]; (4) a particle above a substrate illuminated by a plane wave [13–15].

### B. Force calculation formalism

General expression of the force acting on a dipole particle is [16]

$$F_i = \frac{1}{2} \text{Re}(\alpha \mathbf{E}^{tot} \cdot \partial_i \mathbf{E}^{tot*}), \quad i = x, y, z. \tag{S1}$$

Here we have a scalar product of the electric field vector $\mathbf{E}^{tot}$ and its derivative $\partial_i \mathbf{E}^{tot}$ along one of Cartesian coordinates, i.e. $\mathbf{E}^{tot} \cdot \partial_i \mathbf{E}^{tot} = E_x^{tot} \partial_i E_x^{tot} + E_y^{tot} \partial_i E_y^{tot} + E_z^{tot} \partial_i E_z^{tot}$. Bold capital letters stays for vector fields, capital letters with a subscript denote Cartesian components (scalar) of a vector. $\mathbf{E}^{tot}$ is the total external field at the particle location. The particle radiation self-action is included in the effective particle polarizability α obtained through the electrostatic polarizability $\alpha_{ES}$ [16]:

$$\alpha = \frac{\alpha_{ES}}{1 - i \frac{k_1^3}{6\pi\varepsilon_0} \alpha_{ES}}, \qquad \alpha_{ES} = 4\pi R^3 \varepsilon_0 \frac{\varepsilon - \varepsilon_1}{\varepsilon + 2\varepsilon_1}, \tag{S2}$$

where $\varepsilon_0$ and $\varepsilon_1$ are permittivities of vacuum and the upper half-space respectively, $\varepsilon$ and R are particle permittivity and radius, $\mathbf{k}_1$ is the wave-vector in the upper half-space.

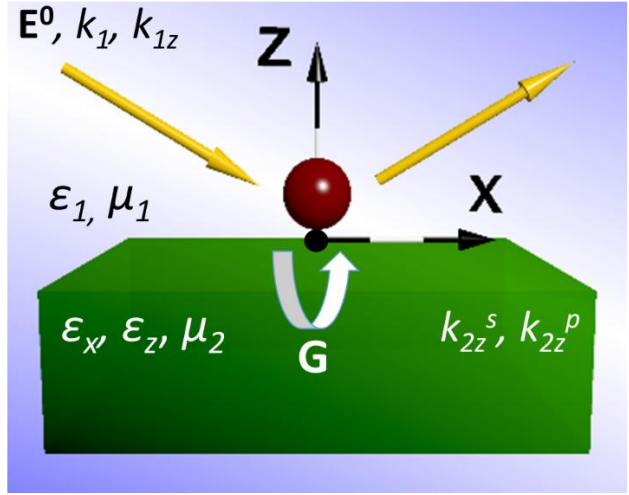

**Figure S1.** A particle of the radius R and the permittivity $\varepsilon$ is located at the distance z above an anisotropic substrate with the permittivity components $\varepsilon_x = \varepsilon_y$, $\varepsilon_z$. The permittivity and permeability of upper half-space are $\varepsilon_1$, $\mu_1$. The origin of coordinates lies at the substrate-air interface.

If the particle is located above a substrate and illuminated by a plane wave, there are two channels constituting the field $\mathbf{E}^{tot}$ at the particle location – plane wave $\mathbf{E}^0$ (incident and reflected from the substrate) and the field scattered by the particle and reflected from substrate $\mathbf{E}^{sc,r}$. With Green's function of the reflected field:

$$\vec{\mathbf{G}} = \begin{bmatrix} G_{xx} & G_{xy} & G_{xz} \\ G_{yx} & G_{yy} & G_{yz} \\ G_{zx} & G_{zy} & G_{zz} \end{bmatrix}, \qquad \partial_i \vec{\mathbf{G}} = \begin{bmatrix} \partial_i G_{xx} & \partial_i G_{xy} & \partial_i G_{xz} \\ \partial_i G_{yx} & \partial_i G_{yy} & \partial_i G_{yz} \\ \partial_i G_{zx} & \partial_i G_{zy} & \partial_i G_{zz} \end{bmatrix}, \quad i = x, y, z, \tag{S3}$$

the field $\mathbf{E}^{sc,r}$ can be obtained as

$$\mathbf{E}^{sc,r} = \omega^2 \mu_0 \mu_1 \vec{\mathbf{G}} \alpha \mathbf{E}^{tot} \tag{S4}$$

where $\alpha \mathbf{E}^{tot} = \mathbf{p}$ is just an induced dipole moment of the particle. Using this expression and writing the total field $\mathbf{E}^{tot} = \mathbf{E}^0 + \mathbf{E}^{sc,r}$ at the particle location

$$\mathbf{E}^{tot} = \mathbf{E}^0 + \omega^2 \mu_0 \mu_1 \vec{\mathbf{G}} \alpha \mathbf{E}^{tot}, \tag{S5}$$

we can solve this equation self-consistently for the total field $\mathbf{E}^{tot}$ in the particle-substrate system taking into account multiple rescattering. With Green's function $\overset{\leftrightarrow}{\mathbf{G}}$ of the reflected field being diagonal at the dipole location, the components of the total field are given by

$$E_i^{tot} = E_i^0 (1 - \omega^2 \mu_0 \mu_1 G_{ii} \alpha)^{-1}, \quad i = x, y, z, \tag{S6}$$

where $\mu_0$ and $\mu_1$ are permeabilities of vacuum and upper half-space, $\omega$ is angular light frequency.

For evaluation of the total field $\mathbf{E}^{tot}$, Eq. S5-S6, Green's function is required. The components of $\overset{\leftrightarrow}{\mathbf{G}}$ (relevant for $p$-polarized incident wave) diagonal at the dipole location $\mathbf{r}_0 = (0,0,z)$ are [16]:

$$G_{xx} = \frac{i}{8\pi} \int_0^\infty k_\rho \left( \frac{r_s}{k_{z_1}} - \frac{r_p k_{z_1}}{k_1^2} \right) \mathrm{e}^{-2ik_{z_1} z} \, \mathrm{d}k_\rho, \quad G_{zz} = \frac{i}{4\pi k_1^2} \int_0^\infty \frac{r_p k_\rho^3}{k_{z_1}} \mathrm{e}^{-2ik_{z_1} z} \, \mathrm{d}k_\rho,$$

$$r_s = \frac{k_{1z} - k_{2z}^s}{k_{1z} + k_{2z}^s}, \quad k_{2z}^s = (\varepsilon_x k_0^2 - k_x^2)^{0.5}. \tag{S7}$$

Here $r_s$ is a reflection coefficient of $s$-polarized plane wave, $k_0$ is the vacuum wavenumber, $k_{1z}$ is $z$-component of the wavevector in the upper half-space, $k_{2z}^s$ is $z$-component of the wavevector of $s$-polarized wave in an anisotropic substrate, $\varepsilon_x$, $\varepsilon_z$ are permittivity components of the substrate, i.e. an uniaxial crystal with the axis pointing normal to the crystal interface with air. The total field can now be found with Eqs. S6, S7 and incident field.

Finally, to obtain the final expression for the optical force, we need to differentiate Eq. S5 to find the field derivative. Field derivative relevant to $F_x$ is:

$$\partial_x \mathbf{E}^{tot} = \partial_x \mathbf{E}^0 + \omega^2 \mu_0 \mu_1 \partial_x \overset{\leftrightarrow}{\mathbf{G}} \alpha \mathbf{E}^{tot}. \tag{S8}$$

The only nonzero components of $\partial_x \overset{\leftrightarrow}{\mathbf{G}}$ are $\partial_x G_{xz} = -\partial_x G_{zx}$ which are given by Eq. 4. Then

$$\partial_x \mathbf{E}^{tot} = \begin{Bmatrix} \partial_x E_x^0 + \omega^2 \mu_0 \mu_1 \partial_x G_{xz} \alpha E_z^{tot} \\ 0 \\ \partial_x E_z^0 - \omega^2 \mu_0 \mu_1 \partial_x G_{xz} \alpha E_x^{tot} \end{Bmatrix}. \tag{S9}$$

With Eq. S9 the final expression for the optical force – Eq. 1 from the main text - is obtained.

For small distances from a substrate the imaginary part of Green's function can be approximated as

$$\text{Im}\left(\partial_x G_{xz}\right) = \frac{1}{8\pi k_1^2} \text{Im}\left(\int_0^\infty r_p k_\rho^3 e^{2ik_{1z}z} \, dk_\rho\right) \approx \frac{1}{8\pi k_1^2} \int_0^\infty \text{Im}(r_p) k_\rho^3 \cos(2k_{1z}z) \, dk_\rho \,. \qquad \text{(S10)}$$

### C. Green's function and excitation conditions

According to Eq. 1 a force enhancement is expected, when the term $\text{Im}(E_x E_z^*) \text{Im}(\partial_x G_{xz})$ is maximized. By multiplying $\text{Im}(\partial_x G_{xz})$ and $\text{Im}(E_x E_z^*)$, depicted in Fig. S1, one can see the direct correspondence to the force, which appears in Fig. 2. In line with Green's function dependence, force peak emerges only at the SPP resonance in the III quarter, while, in contrast, the negative force is achievable with a broader set of $\varepsilon_x'$, $\varepsilon_z'$ parameters in the II quadrant. In the IV quadrant, the excitation conditions give negligible $\text{Im}(E_x E_z^*)$, prohibiting the force enhancement, despite the large values of $\text{Im}(\partial_x G_{xz})$.

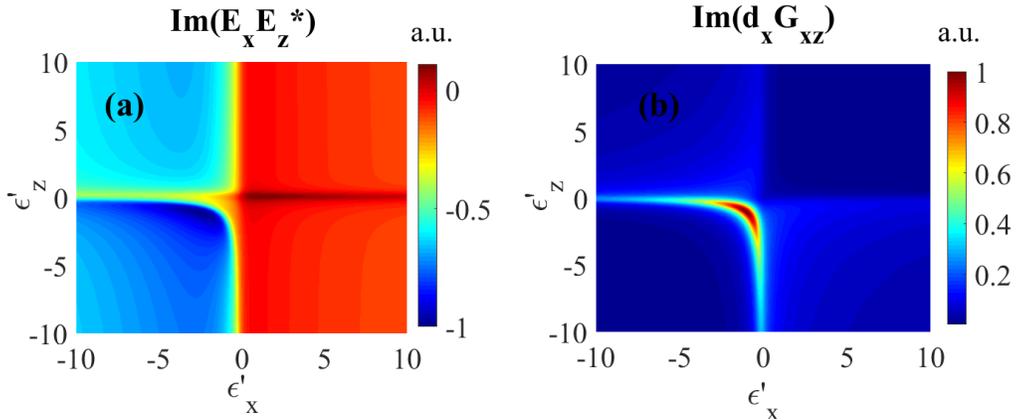

**Figure S1.** (a) The product of the field components at the particle illuminated by a plane wave, incident at 35º and (b) Green's function component $\text{Im}(\partial_x G_{xz})$ for the anisotropic substrate with $\varepsilon_x''$, $\varepsilon_z'' = 0.3$ at coordinate $z = 15$ nm.

In Fig. 2 (a) the highest $F_x$ is offered in the III quadrant. However, highly anisotropic metals rarely exist in nature, and such artificial materials are difficult to fabricate. As the material losses increase, anisotropic and isotropic metals in the III quadrant provide the same level of enhancement. For isotropic Ag interface ($\varepsilon'' < 0.5$), the plasmon resonance takes place at 340nm wavelength. Metals, suitable for the operation at visible and infrared, like Au, resonate at 510 nm, and have much higher absorption. Consequently, weaker pulling forces can be achieved. Increasing the imaginary part of $\varepsilon_x''$, $\varepsilon_z''$ up to 3 and higher (the level of values, present in natural materials, e.g. Au, Cu, Al) smears out the resonance. As the result, the force enhancement, provided by surface plasmons (III quarter), become comparable with the contribution of the hyperbolic modes (II quarter).

### D. Applicability of the effective medium theory to the optical forces calculations

Straightforward investigation of the transversal force by variation of the metal fraction in the multilayer structure is shown in Fig. S2(a). The force is normalized to the intensity of an incident beam. The sharp peak around 340 nm corresponds to the surface plasmon resonance. At the visible range (relevant to experimental realizations) one can see that the multilayer substrate provides noticeable force enhancement at practically any wavelength. Figure S2(b) presents the results, obtained with the effective medium approximation.

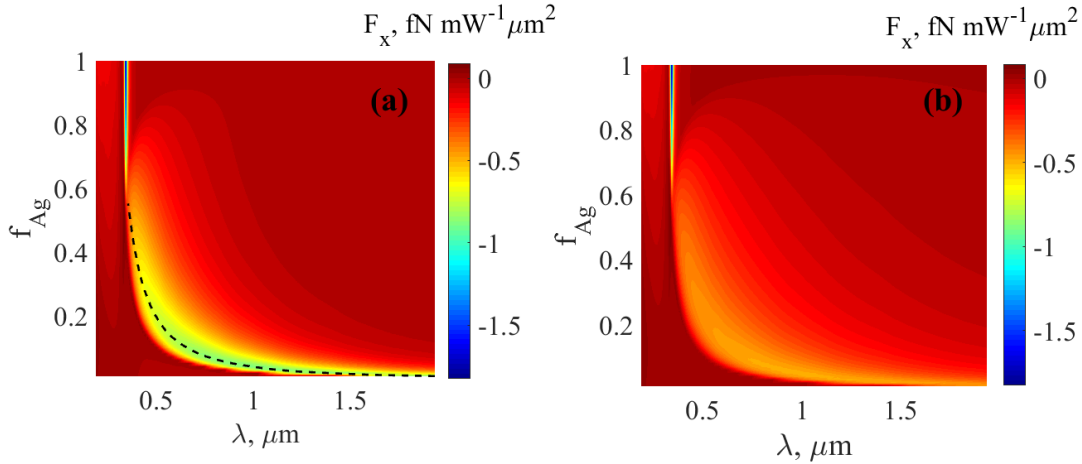

**Figure S2.** Transversal force $F_x$ acting on the particle ($R = 15$ nm, $\varepsilon = 3$, $z = 15$ nm) above (a) the semi-infinite multilayer with the period $P = 5$ nm or (b) corresponding homogenized material, as the functions of the wavelength and the metal filling factor. The multilayer is formed by Ag and a porous polymer with the top metallic layer and illuminated by a plane wave incident at 35º.

Figure S3(a) shows the peak force values (blue envelope) for multilayers with metal on the top ($P$=8 nm) along with the corresponding spectra. Figure 3S(b) compares how the same substrate would behave if the topmost layer was dielectric (light blue line). The force drops about an order of magnitude in the case of the top dielectric layer. Metal layer thickness $d$ is plotted in the inset in Fig. S3(b) for each wavelength.

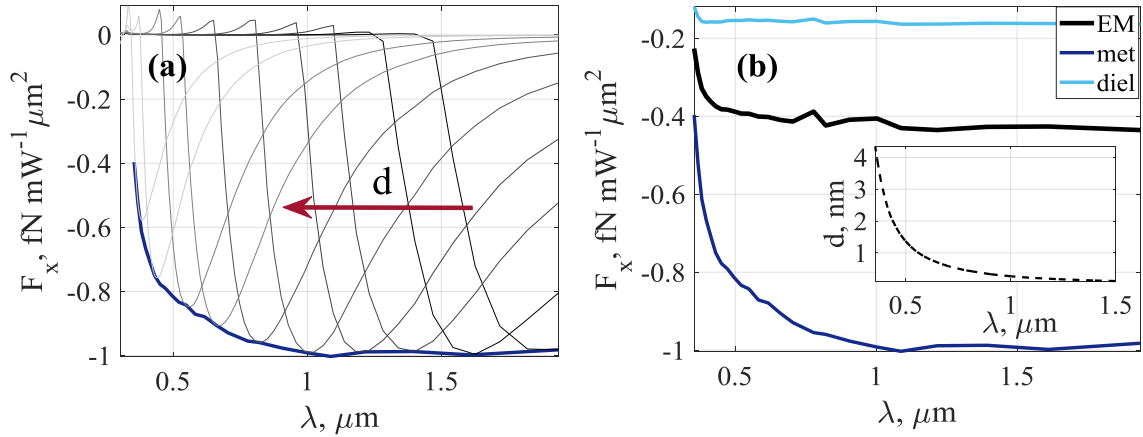

**Figure S3.** Transversal force $F_x$ acting on the particle ($R$ = 15 nm, $\varepsilon$ = 3, $z$ = 15 nm) lying on the Ag-polymer multilayer (terminating with metal) with the period $P$ = 8 nm and illuminated with a plane wave incident at 35°. (a) $F_x$ spectra for the multilayers optimized to give maximal force at each wavelength. Dark blue line (given by dashed black line in Fig. S2 (a)) is the envelope giving the peak enhancement. (b) $F_x$ envelope from (a) and analogous envelope given by the effective medium approximation (black line) and by the multilayer with the dielectric on the top (light blue line). Inset shows the metal layer thickness $d$ in the unit cell for each wavelength.

### E. Optimization of the multilayers for achieving negative optical forces

Figure S4 compares $F_x$ for different types of metal constituting multilayer, Ag, Au or Cu. For each wavelength, the multilayer is optimized to obtain the highest force values. Silver shows higher values along the spectra, especially at shorter wavelengths, where there is a drop for Au and Cu. For each wavelength the silver layer thickness $d$ giving maximal force enhancement from (a) is plotted in Fig. S4(b). The optimal $P$ for all metals is around 8-10 nm. For the smallest period the metal layer becomes thicker, what is favourable for a fabrication, while for the largest period $P$ = 12 nm $d$ decreases. At IR the vanishing metal thickness suggests that graphene could be used instead of a metal, as it was done for the $F_z$ component in [6].

Figure S4(c,d) shows the effective dielectric permittivities $\varepsilon_x$ and $\varepsilon_z$. The imaginary part of the effective permittivity in multilayers is much lower than in the case of a pure metal. As it is evident, $\varepsilon_x'$ is near-zero and negative while $\varepsilon_z'$ is positive being from 1 to 2, which corresponds to II quadrant in Fig. 2.

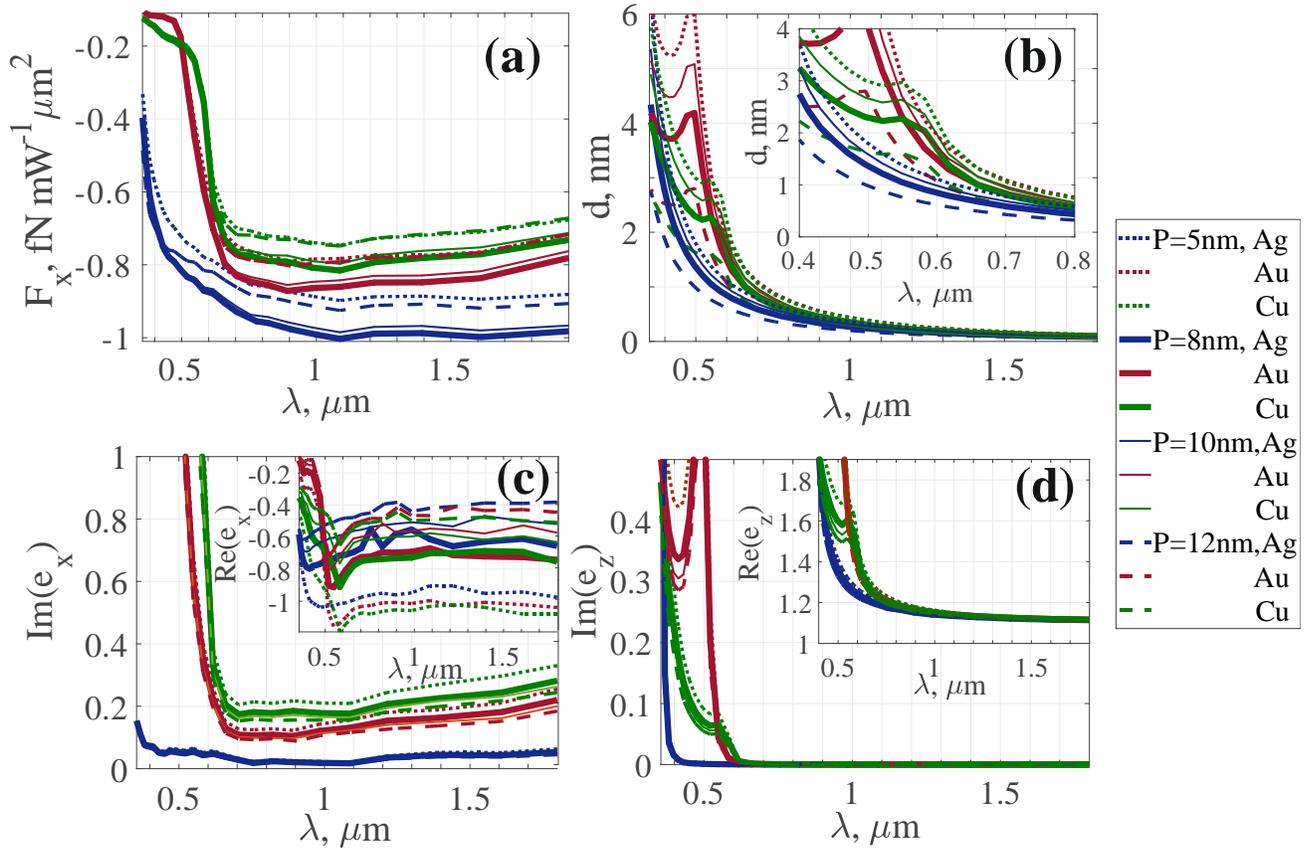

**Figure S4.** (a) Transversal force $F_x$ (optimized for each wavelength) for the particle ($R = 15$ nm, $\varepsilon = 3$, $z = 15$ nm) touching multilayer with the unit cell comprising porous polymer and metal: Ag (blue color), Au (red color) or Cu (green color) with different periods (see the legend). The outer material is the metal; a plane wave is incident at 35°. The thickness of the metal layer (b) and the effective dielectric permittivities $\varepsilon_x$ (c) and $\varepsilon_z$ (d) corresponding to the optimized force values from (a).

## F. Particles suspended in water

In experimental realizations water environment is widely employed. If the upper half-space is filled with water instead of air, the effective particle polarizability given by Eq. S2 becomes smaller, and the optical force decreases. Additionally, the hyperbolic modes affect the optical force less according to the decrease of the imaginary part of $r_p^\infty = (\sqrt{\varepsilon_x \varepsilon_z} - \varepsilon_1)/(\sqrt{\varepsilon_x \varepsilon_z} + \varepsilon_1)$. Figure S5 shows that in water environment optically denser particles can be used to achieve sufficient force magnitudes.

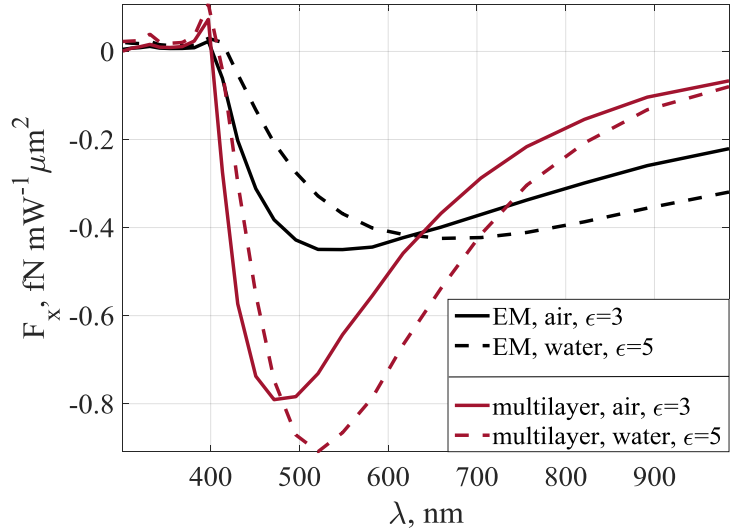

**Figure S5.** Transversal force $F_x$ acting on the particles of different permittivities, $\varepsilon = 3$ or $\varepsilon = 5$, suspended in air or water. Red lines correspond to the semi-infinite Ag-polymer multilayer ($f_1 = 0.2$) substrate with a metal layer on top and the period $P = 8$ nm. Black lines are obtained within the effective medium approach. Other parameters are: $R = 15$ nm, $z = 15$ nm, the structure is illuminated by a plane wave incident at 35°.